\documentclass[usenatbib]{mn2e}

\usepackage{graphicx,hyperref,amsmath,bm}
\hypersetup{colorlinks,linkcolor=blue, citecolor=blue}

\begin{document}

\title[BH--NS Mergers In Globular Clusters]{Black Hole--Neutron Star Mergers in Globular Clusters}
\author[D. Clausen et al.]{Drew Clausen,$^1$\thanks{Email: dclausen@astro.psu.edu} Steinn Sigurdsson,$^1$ David F. Chernoff $^2$\\
$^{1}$Department of Astronomy \& Astrophysics, The Pennsylvania State University, 525 Davey Lab, University Park, PA 16802\\
$^{2}$Department of Astronomy, Cornell University, 610 Space Sciences Building, Cornell University, Ithaca, NY 14853}

\date{Accepted 2012 October 29}
\maketitle

\begin{abstract}
We model the formation of black hole--neutron star (BH--NS) binaries via dynamical interactions in globular clusters.  We find that in dense, massive clusters, 16--61\% of the BH--NS binaries formed by interactions with {existing} BH binaries will undergo mergers driven by the emission of gravitational radiation.  If the BHs are retained by the cluster after merging with a NS, the BHs acquire subsequent NS companions and undergo several mergers.  Thus, the merger rate depends critically upon whether or not the BH is retained by the cluster after the merger.  Results from numerical relativity suggest that kick imparted to a $\sim 7~\rm{M_{\sun}}$ BH after it merges with a NS will greatly exceed the cluster's escape velocity.  In this case, the models suggest that the majority of BH--NS mergers in globular clusters occur within 4 Gyrs of the cluster's formation and would be unobservable by Advanced LIGO.  For more massive BHs, on the other hand, the post merger kick is suppressed and the BH is retained. Models with 35~${\rm M_{\sun}}$ BHs predict Advanced LIGO detection rates in the range $0.04 - 0.7~{\rm yr^{-1}}$.  On the pessimistic end of this range, BH--NS mergers resulting from binary--single star interactions in globular clusters could account for an interesting fraction of all BH--NS mergers.  On the optimistic end, this channel may dominate the rate of detectable BH-NS mergers.                       
\end{abstract}

\begin{keywords}
black hole physics -- gravitational waves --  stars: neutron -- globular clusters: general
\end{keywords}

%%%%%SECTION 1%%%%%
\section{Introduction}
Surveys of the Milky Way globular cluster (GC) system have discovered 13 low mass X-ray binaries \citep{Deutsch:2000,Sidoli:2001} and 143 pulsars, many of which are recycled, millisecond pulsars (MSPs) and/or members of exotic binaries \citep{Lynch:2011,Freire:2007,Hessels:2006,Ransom:2005,DAmico:2001}.  These observations reveal that 1) some fraction of the neutron stars (NSs) formed in GCs are retained despite their natal kicks, and 2) multibody interactions in the cores of GCs greatly enhance the formation rate of binary systems that contain a NS, resulting in an over abundance of low mass X-ray binaries and MSPs relative to the field \citep{Katz:1975,Clark:1975,Verbunt:1987}.  It has long been recognised that binary--single star interactions between NSs and primordial binaries can result in the NS exchanging into the binary by ejecting one member and becoming bound to the other \citep{Hills:1976,Hut:1983a,Sigurdsson:1995,Ivanova:2008}.  Such binary--single star interactions could also result in the formation of black hole (BH)--NS binaries \citep{Sigurdsson:2003,Devecchi:2007}, which are expected to produce gravitational waves detectable by the ground-based interferometer LIGO if they coalesce \citep{Abbott:2009}.  Observations of a BH--NS merger would provide a test of General Relativity in the dynamical strong-field regime and could probe the NS equation of state \citep{Kyutoku:2009,Duez:2010}.  Additionally, BH--NS mergers are also a likely progenitor of some short gamma-ray bursts \citep{Nakar:2007}.      

Whether or not dynamical interactions in GCs can efficiently produce BH--NS binaries is unclear.  Models suggest that the $\sim10-300$ BHs that form within a GC will rapidly sink to the core of the cluster and interact with, and eject one another, severely depleting the cluster's BH population within 1 Gyr \citep{Kulkarni:1993,Sigurdsson:1993,OLeary:2006,Moody:2009,Banerjee:2010, Aarseth:2012}.  Even though most of GC's BHs are ejected early in its evolution, there are models \citep{Miller:2002a} and mounting observational evidence that show some GCs retain at least one BH.  Several extragalactic GCs harbour highly variable X-ray sources with luminosities $\ge 10^{39} {\rm\;erg\;s^{-1}}$, well in excess of the Eddington luminosity of an accreting NS, that are strong candidates for BHs in GCs \citep{Maccarone:2011}.   Even if a GC only retains a single BH, it may be possible for this BH to merge with multiple NSs if the BH exchanges into another NS-containing binary after each merger.  

What is the maximum rate for such successive mergers?  If we assume that the BH and NS merge instantaneously, then the maximum merger rate is equal to the rate at which the BH can exchange into a NS-containing binary, $R_{\rm ex} =n_{\rm NS}\sigma_{\rm ex} v_{\rm rel}$, where $n_{\rm NS}$ is the density of NS-binaries, $\sigma_{\rm ex}$ is the cross section for an exchange, and $v_{\rm rel}$ is the relative velocity of the BH and the binary.  \citet{Ivanova:2008} found that GCs tend to retain $\sim 220$ NS per $2\times10^{5} \rm{M_{\sun}}$.  Due to mass segregation, NSs will be concentrated {near} the centre of a GC, and the central NS fraction, $f_{\rm NS}$, can be as high as 0.1.  Assuming that all of the NSs at the centre of the cluster are in a binary and that the collision cross section is dominated by gravitational focusing, we can approximate the maximum BH--NS merger rate as 
\begin{equation}
\begin{split}
R_{\rm ex} \sim 2 \times 10^{-10}~{\rm yr^{-1}} \left(\frac{f_{\rm NS}}{0.1}\right)\left(\frac{n}{10^{5} {\rm~pc^{-3}}}\right)\left(\frac{a}{1 {\rm~au}}\right)\\ \times  \left(\frac{M_{\rm BH}}{1~{\rm M_{\sun}}}\right) \left(\frac{\bar{v}_{m}}{10~{\rm km~s^{-1}}}\right)^{-1}.
\end{split}
\end{equation}         
Here, $n$ is the total central stellar density, $a$ is the typical semi-major axis of a NS binary, $M_{\rm BH}$ is the BH mass, and $\bar{v}_{m}$ is the mean central velocity dispersion.  

In this paper we present detailed models that investigate the rate at which the BHs and NSs retained by a GC will form binaries that merge via the emission of gravitational radiation and the corresponding detection rate for LIGO.  We will begin by discussing the methods used in our simulations in \autoref{sec:meth}.  In \autoref{sec:sms}, we will describe the results of simulations that only allowed BHs to merge with a single NS, and in \autoref{sec:multi} we will describe how the results change when the BH is allowed to under go successive mergers as sketched above.  We will conclude in \autoref{sec:disc} by computing the LIGO detection rate and comparing our work with previous calculations.    

%%%%%SECTION 2%%%%%
\section{Method}
\label{sec:meth}
We modelled the formation of BH--NS binaries through binary--single star interactions using the method presented in \citet{Sigurdsson:1995} and \citet{Mapelli:2005}.  For each simulation, an ensemble of binaries was evolved in a static background cluster.  The background clusters were multi-mass King models, with the stars of each mass group $\alpha$ following the isotropic distribution function 
\begin{equation}
	f_{\alpha}(\epsilon) =
	\begin{cases}
	 \frac{n_{0_{\alpha}}}{(2\pi v_{m_{\alpha}}^{2})^{3/2}}[e^{\epsilon/\sigma_{\alpha}^{2}}-1] &\epsilon > 0\\
	 0& \epsilon \le 0\\
	\end{cases}
	,
	\label{eqn:dist}
\end{equation}
where $v_{m_{\alpha}}$ is the mass group's core velocity dispersion, $n_{0_{\alpha}}$ is a normalising constant, and $\epsilon = \Psi(r) - v_{*}^{2}/2$ is the relative energy per unit mass.  Here, $v_{*}$ is the velocity and $\Psi(r)$ is the relative potential given by $\Phi(r_{t}) - \Phi(r)$, where $\Phi(r)$ is the gravitational potential {with respect to infinity} at a distance $r$ from the cluster centre and $r_{t}$ is the cluster's tidal radius. To explore how the BH--NS merger rate is impacted by the structural properties of the GC, we varied the central stellar number density, $n_{c}$, the number density weighted mean central velocity dispersion, $\bar{v}_{m}$, and the King parameter $W_{0} \equiv \Psi(0)/\bar{v}_{m}^{2}$ to generate clusters of different mass and concentration.  The values of $n_{c}$, $\bar{v}_{m}$, and $W_{0}$ were chosen to create a set of background clusters with core radii $r_{c}$ and concentrations $c = \log (r_{t}/r_{c}) $ similar to observed clusters.  Furthermore, these clusters probe a range of core interaction rates $\Gamma \propto n_{c}^{1.5}r_{c}^{2}$, a quantity that is known to correlate with another product of dynamical interactions, the number of X-ray binaries present in a cluster \citep{Pooley:2003}.  We list the values used in our models in \autoref{tab:gcparms}.      

For each cluster, we used the initial stellar mass function
\begin{equation}
 \xi(m) \propto 
 \begin{cases}
 m^{-0.3} &m < 0.35\;{\rm M_{\odot}} \\
 m^{-2.35} & m > 0.35\;{\rm M_{\odot}} 
\end{cases}
\end{equation} 
\citep{Kroupa:2001}, and binned the evolved population into 10 mass groups.  The main sequence (MS) turnoff mass, $m_{\rm to}$, was set to $0.85\; {\rm M_{\odot}}$ and stars with initial mass above the turnoff mass were assumed to be completely evolved.  The evolved stars fell into one of three groups, those with initial mass $m_{i}< 8 {\rm\;M_{\odot}}$ evolved into white dwarfs (WDs) with masses given by $m_{WD} = 0.38+0.12\;m_{i}$ \citep{Catalan:2008}. Stars with initial mass in the range $8 {\;\rm M_{\odot}} \le m_{i} < 25 {\;\rm M_{\odot}}$ were assumed to form $1.4 {\rm\;M_{\odot}}$ NSs and those with initial mass $> 25{\;\rm M_{\odot}}$ formed BHs. 

We performed calculations using two different values for the BH mass.  {Recent statistical analyses of low mass X-ray binaries in the Galaxy have found that the masses of BHs are narrowly distributed around $7~{\rm M_{\odot}}$ \citep{Ozel:2010,Farr:2011}.  Motivated by this,} we have run models in which all BHs have a mass of $7~{\rm M_{\odot}}$.  However, the progenitors of the field BHs considered in the above studies were stars with much higher metallicity than those found in GCs.  Lower metallicity stars have reduced mass loss rates and could evolve to BHs with masses as high as $80~{\rm M_{\odot}}$ \citep{Fryer:2001,Fryer:2002,Belczynski:2010}.  There is also observational evidence that BHs in GCs are more massive than those in the Galaxy. Properties of the ultraluminous X-ray sources observed in GCs associated with NGC 4472, M31, and NGC 1399 are consistent the presence of a BH of $\ga 30~{\rm M_{\odot}}$ \citep{Maccarone:2007,Barnard:2011,Irwin:2010,Clausen:2012}.  Accordingly, we also ran models with $35~{\rm M_{\odot}}$ BHs to allow for the possibility that GC BHs are more massive than those in the field.  

We assumed that the clusters retained 20\% of the NSs that were formed within them.   We further assumed that nearly all BHs formed in the cluster were promptly ejected by self interaction, leaving 0, 1, or 2 BHs \citep{Kulkarni:1993,Sigurdsson:1993,OLeary:2006,Moody:2009,Banerjee:2010,Aarseth:2012}. We mimic this by truncating the high mass end of the initial stellar mass range so there is a single BH in the GC model.  We are, therefore, modelling the late evolution of the cluster, and hence recent times, so the merger rates calculated here apply to the local universe.   {In all of our simulations, we assumed that the remaining BH was in a binary.  The properties of such a binary will have undergone extensive modification driven by both stellar and dynamical evolution.  Consequentially, the initial configurations assigned to the BH binaries in our models represent the result of complicated processes that we do not attempt model.}
\begin{table}
\centering
 \caption{Background Globular Cluster Models Parameters\label{tab:gcparms}}
  \begin{tabular}{@{}cccccc@{}}
	\hline
	Model&$n_{c}$& $\bar{v}_{m}$&$W$& $M$ & $c$\\
	Name&$({\rm pc^{-3}})$&$({\rm km\;s^{-1}})$&& ($\rm{M_{\odot}}$)& \\
	\hline
	A&$1\times10^{4}$&6&6&$1.0\times10^{5}$&1.20\\
	B&$1\times10^{5}$&10&10&$5.2\times10^{5}$&1.71\\
	C&$5\times10^{5}$&11&13&$7.2\times10^{5}$&1.93\\
	D&$1\times10^{6}$&12&15&$1.1\times10^{6}$&2.06\\
	\hline
 \end{tabular}
 \end{table}

%%%%%SECTION 2.1%%%%%
\subsection{Dynamics}
\label{sec:norm}
For each simulation, we evolved 2000 binaries, one at a time, in one of the background clusters described above.  Each binary contained a BH and a companion drawn at random from the cluster's evolved mass distribution.  The initial semi-major axis, $a$, was drawn from a distribution that is flat in $\log a$ between $10^{-3}$ au and $a_{\rm max}$ \citep{Abt:1983}.  {We used a different value of $a_{\rm max}$ for each background cluster, namely 100 au, 33 au, 15 au, and 10 au for clusters A, B, C, and D, respectively.  These values of $a_{\rm max}$ were chosen to cover the range of possibilities that might result from the complicated evolutionary histories of the BH binaries mentioned above.}    The initial eccentricity was drawn from $f(e) \propto 2e$ \citep{Duquennoy:1991}.  

Each binary was initially placed in the cluster core, with its position selected from the radial density distribution of the third most massive mass group.  The results are
insensitive to this somewhat arbitrary choice because the initial position and velocity of the binary are rapidly forgotten.  The binaries were evolved in the cluster potential with dynamical friction and random kicks to account for scattering by individual stars. At each point along the binary's trajectory, we computed the probability that it would experience a strong encounter with a single star using
\begin{equation}
 	P_{\rm enc} = \Delta t \sum_{\alpha} \int n_{\alpha}(r)\sigma(\bm v,\bm v_{*}) |\bm v- \bm v_{*}|f_{\alpha}(\bm v_{*})d^{3}{\bm v_{*}} ,   
	\label{eqn:encprob}     
\end{equation}
where the subscripts $\alpha$ correspond to the mass groups, $\Delta t$ is the time step, $n$ is the number density, $f({\bm v_{*}})$ is the velocity distribution of single stars given in \autoref{eqn:dist}, ${\bm v}$ and ${\bm v_{*}}$ are the velocity of the binary and a single star, respectively, and  $\sigma$ is the encounter cross section.  Throughout this paper we are interested in encounters that alter the energy, angular momentum or components of the binary.  Previous studies have shown that the bulk of these changes involve a {background} star approaching the binary centre of mass closer than a few times the binary's physical size and conversely that numerous, more distant encounters have relatively small effect.  We choose the critical pericentre distance for simulated encounters to be $p = a [4 + 0.6(1+e)]$ where $a$ is the binary semi-major axis and $e$ is its eccentricity \citep{Hut:1983a}.  The corresponding cross section for such encounters is
\begin{equation}
	\sigma(\bm v, \bm v_{*}) = \pi p^{2}  + \frac{2 \pi G (m_{\rm bin} + m_{\alpha}) p}{|\bm v- \bm v_{*}|^2},
\end{equation}
where $m_{\rm bin}$ is the mass of the binary. An encounter took place if  $P_{\rm enc}$ was greater than a random number drawn from a uniform distribution between 0 and 1.  {Given that an encounter occurred, we randomly chose which particular mass group was involved based on the fractional encounter probabilities of all the groups.}   Then, the three-body interaction between the stars was integrated explicitly as described in \citet{Sigurdsson:1993a}.  There were many possible outcomes to the three-body encounter, including mergers, exchanges, and disruption of the binary, but if a binary still existed at the end of the interaction, then we continued to evolve the binary in the cluster.  

We performed two types of simulations.  In the first class of simulations we were only concerned with the fate of the {initial} BH binary, while in the second class we sought to determine the ultimate fate of the BH.  In the former set, a run continued until either the initial binary was disrupted, merged, or ejected, or the run had covered $10^{10}$ yrs. In the latter set of models, we continued a run when the merger or disruption of a binary resulted in an isolated BH (see \autoref{singles}).  These runs ended only after the BH was ejected from the cluster or the maximum run time of $10^{10}$ yrs  was reached. In both simulation types, a small fraction of the runs $(\la 5\%)$ were aborted because integration of a three-body interaction exceeded the maximum number of allowed steps, $8 \times 10^{7}$.       

%%%%%SECTION 2.2%%%%%
\subsection{Gravitational radiation effects on the evolution BH--NS binaries}
\label{GR}
Between interactions the eccentricity and semi-major axis of each BH-NS system were decreased to account to account for the emission of gravitational radiation using the relations given in \citet{Peters:1964}.  For the simulations in which we used a BH mass $M_{\rm BH} = 7~{\rm M_{\odot}}$, we stopped the calculation when a BH--NS binary coalesced.  Numerical relativity simulations have found that the remnant formed by a BH-NS merger at this mass ratio will receive a kick from anisotropic emission of  gravitational radiation that exceeds the GC's escape velocity \citep{Shibata:2009,Etienne:2009,Foucart:2011}.  In the models where we used $M_{\rm BH} = 35~{\rm M_{\odot}}$, the post-merger recoil velocity was suppressed because of the much smaller mass ratio ($q = M_{\rm NS}/M_{BH}$).  For the case of non-spinning point masses, the merger of a $1.4~\rm{M_{\odot}}$ NS and a $35~{\rm M_{\odot}}$ BH would result in a recoil velocity of $14.6~{\rm km~s^{-1}}$, well below the escape velocity for the clusters models that we considered \citep{Gonzalez:2007}.  However, we allowed for spinning BHs and NSs, and computed the recoil velocity using the parameterisation given in \citet[however, see \citealt{Hirata:2011} for discussion of a resonant recoil, not included in these parameterisation, that can dominate when the BH spin is nearly maximal and $q\ll1$]{Campanelli:2007}.  At the beginning of each run, the BH spin parameter was randomly selected from a uniform distribution between 0 and 1.  When a BH--NS binary merged, the NS spin was randomly selected from a uniform distribution between 0 and 0.7 (a spin parameter of 0.7 corresponds to a NS spinning at the break up frequency \citealt{Miller:2011}) and the two spins were randomly oriented with respect to one another and to the orbital plane.  The result was a range of kick velocities ($\sim 10-100~{\rm km~s^{-1}}$), some of which were low enough that the single BH was retained by the cluster after the merger.   

%%%%%SECTION 2.3%%%%%
\subsection{Formation of a new binary from a single BH by stellar interactions}
\label{singles}
In some of our models, when the BH lost its binary companion, either through an interaction with a {background} star or as the result of a gravitational radiation driven merger with a NS, we continued to evolve the now single BH in the cluster.  As the BH moved through the cluster, we sought to determine the single BH's next binary companion by generating  binaries with a semi-major axis $a$, an eccentricity $e$, a primary of mass $m_{1}$, and a secondary of mass $m_{2}$.  The semi-major axis and eccentricity were drawn from the same distributions as those of the {initial} BH binaries. The only difference is that we have chosen $a_{\rm max} = 1$ au for these systems to ensure that the binaries are hard.  Our choices for $m_{1}$ and $m_{2}$ are described below.  

{We determined whether or not the single BH would interact with the binary by temporarily replacing the secondary with the BH.  We then calculated the probability that this temporary binary would interact with a  background star of mass $m_{2}$ (the secondary), $P_{\rm (1,BH) + 2}$, using \autoref{eqn:encprob}.  Next, $P_{\rm (1,BH) + 2}$ was scaled to $P_{\rm (1,2) + \rm {BH}}$ by comparing the rates of two related exchange encounters, $[(1,\rm BH)+ 2 \rightarrow (2,{\rm BH})+1]$ and $[(1,2)+{\rm BH}\rightarrow (1,{\rm BH})+2]$.  The rate for an arbitrary interaction between a binary (a,b) and single star c can be expressed in the form 
\begin{equation}
R(X) = \frac{G m_{a}m_{b}(m_{a}+m_{b}+m_{c})}{m_{c}(m_{a}+m_{b})} \frac{a_{\rm in} n \tilde{\sigma}(X)}{v}, 
\label{eqn:ratescale}
\end{equation}
where $X$ refers to the type of interaction (e.g., one of the desired exchange reactions), $v$ is the relative velocity of the binary and the single star, $a_{\rm in}$ is the semi-major axis of the binary, $n$ is the density of single stars,  $\tilde{\sigma}(X)$ is a numerically determined, dimensionless cross section, $G$ is the gravitational constant, and $m_{i}$ is the mass of star $i$.  Using \autoref{eqn:ratescale}, we computed the relative rates of the above exchange interactions, assuming that $a_{\rm in}$, $n$, and $v$ were the same in each case. 

The required dimensionless cross sections were computed by  \citet{Sigurdsson:1993a}, who found $\tilde{\sigma}[(1,\rm BH)+ 2 \rightarrow (2,{\rm BH})+1] \sim 2\times \tilde{\sigma}[(1,2)+{\rm BH}\rightarrow (1,{\rm BH})+2]$.  In addition, we made the simplifying approximation that $m_{1} = m_{2} \ll M_{\rm BH}$, because the BH was much more massive than either the primary or the secondary.   This reduced the ratio of the mass factors to $m_{2}/2M_{\rm BH}$.  Combining these, we approximated the probability that the single BH would interact with a binary containing the primary and the secondary as
\begin{equation}
\label{eqn:scale}
P_{\rm (1,2) + BH} \sim \frac{f_{b}}{1-f_{b}} \frac{m_{2}}{M_{\rm BH}} P_{\rm (1,BH) + 2},
\end{equation}
where $f_{b}$ is the binary fraction and relates the densities of single and binary stars.  Note that the right hand side of this expression diverges as $f_{b} \rightarrow 1$.  Thus, for large values of $f_{b}$, use of \autoref{eqn:scale} will overestimate the rate at which a single BH can exchange back into a binary. Furthermore, we have used the relative rates of two {\em exchange} encounters to scale the {\em total} encounter probability. The exchange cross section is a significant fraction of the total interaction cross section, but we will explore the consequences of this choice in \autoref{sec:unc}.  {We chose to compute the encounter probability using the scaling relation given in \autoref{eqn:scale} because the distribution of binaries in globular clusters is poorly understood and this technique offers an efficient and adequate estimation of the rates within the limitations of our code.} When an encounter was deemed to have occurred, we restored the original binary and explicitly integrated the three-body interaction between the binary and the BH, accepting any outcome.  We repeated this process until the BH exchanged into a binary, the BH was ejected from the cluster, or the maximum integration time was reached.  If the BH exchanged into a binary, we continued the simulation as described in \autoref{sec:norm}.}

We explored several different binary populations and values of $f_{b}$.  In one set of models, we drew the primary's mass from the cluster's evolved mass distribution and required that $m_{1} \ge m_{\rm to}$, but we did not chose the secondary's mass, initially.  Instead, we used a binary consisting of  the primary and the BH to compute the scaled probability that a BH would interact with binary containing the primary and a member of any mass group.  If an encounter occurred, then the secondary was chosen to belong to the mass group that interacted with the temporary binary.  We then integrated the three-body interaction as described above.  This method of selecting the secondary resulted in a binary population that is comprised largely of systems that contain a NS.  For the simulations in cluster D, more than 50\% of the binaries that interacted with a single BH contained a NS.  We have labeled the runs that use this binary population OPT, for optimised, because this method produces a binary population that is tuned to interact with the single BH.  Furthermore, because many of the binaries in this population already contain a NS, the single BH can gain a NS companion immediately.

The second binary population that we considered was closely linked to observational constraints.  A number of studies have used the fact unresolved MS binaries appear brighter and redder than single MS stars to investigate binaries in GCs.  The binary fraction can be measured by comparing the number of objects in the offset binary sequence to the number of stars that lie along the MS, with observed values typically found in the range $f_{b} = 0.05 - 0.3$ \citep[e.g.,][]{Romani:1991, Sollima:2007,Davis:2008, Milone:2012}. We generated a population that resembles these observed populations by independently drawing two stars from the cluster's evolved stellar mass distribution and requiring that one of the stars have a mass $m \ge m_{\rm to}$.  The result was a collection of systems that consisted primarily ($\ga 85\%$) of MS-MS binaries.  We have labeled simulations that use this binary population OBS, and considered values for the binary fraction within the observed range: 0.05, 0.1, and 0.2.   

We also considered a population that was motived by detailed models of binary evolution in GCs.  Analytic, Monte Carlo, and N-body models have indicated that over time the hard binary fraction will increase in the cluster core \citep{Sollima:2008,Fregeau:2009,Hurley:2007}. For such models to be consistent with the observations described above, the initial binary fraction in GCs would have to have been quite low.  Alternatively, as \citet{Fregeau:2009} suggested, if many of the binaries in a GC consisted of a MS star and a WD or two WDs, they would be ``hidden'' from observers.  Following on this result, we have run  simulations that use a binary population similar to that described in table 1 of \citet{Fregeau:2009}.  This population consisted of 44\% MS--WD binaries, 32\% WD--WD binaries, 23\% MS--MS binaries, and 1\% binaries containing a NS.  Runs that used this binary population were labeled FIR.  Furthermore, because only 23\% of this binary population would be measured by the observations described above, we have used larger values of $f_{b}$ with this population: 0.2, 0.5, and 0.75.           

%%%%%SECTION 3%%%%%
\section{Results}
In all of our simulations, binary--single star interactions produced BH--NS binaries.  However, the number of BH--NS formed and the properties of these systems were very sensitive to the structure of the GC and the assumed binary population.  Our simulations are summarised in \autoref{tab:runs}.  The initial conditions for each simulation, including  the cluster model, the binary population (where applicable), the binary fraction (where applicable), and $M_{\rm BH}$, are given.  Also listed are $f_{\rm BHNS}$, the fraction of runs that produced a BH--NS binary, $N_{\rm BHNS}$, the total number of BH--NS binaries formed in all 2000 runs, $N_{\rm mrg}$, the number of BH--NS mergers during the simulation, $\bar{t}_{\rm ex}$, the average amount of time taken for a single BH to exchange back into a binary, $N_{\rm max}$, the maximum number of mergers in a single run, and $ \mathcal{R}_{GC}$, the BH-NS merger rate for that model.  We describe the results in detail below, beginning with models that assumed that the BH was ejected from the GC after merging with a NS, followed by descriptions of models with more massive BHs, which allowed for a single BH to merge with multiple NS.
\begin{figure}
  \includegraphics[width=0.5\textwidth]{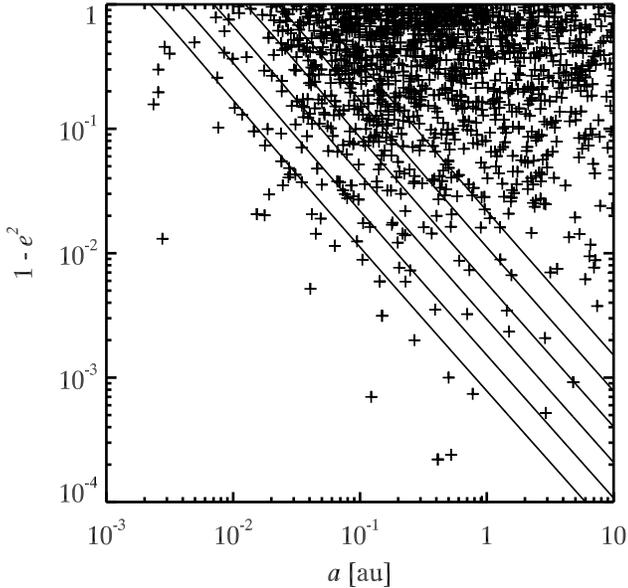}%
  \caption{Initial locations of the BH--NS binaries in the $(1-e^{2}$)--$a$ plane.  The points shown here are from simulations in cluster model D with $M_{\rm BH}= 7~{\rm M_{\odot}}$. $t_{\rm GW_{0}}$ is constant along the solid lines.  From left to right, the lines correspond to $t_{\rm GW_{0}} = 10^{5}, 10^{6}, 10^{7}, 10^{8}, 10^{9}, 10^{10}$ yrs.\label{fig:esqa}}
\end{figure}
\begin{figure}
  \includegraphics[width=0.5\textwidth]{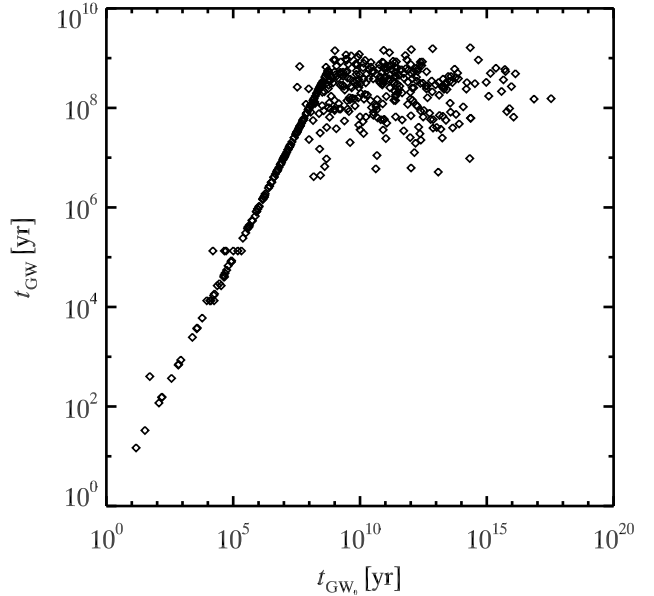}%
  \caption{The actual amount of time it took a BH--NS binary to merge $t_{\rm GW}$ vs. the amount of time expected for the BH--NS binary to merge, given the binary's initial configuration, $t_{\rm GW_{0}}$ for the BH--NS binaries that merge in a simulation of cluster model D with $M_{\rm BH}= 7~{\rm M_{\odot}}$.  Note that many systems formed with  $t_{\rm GW_{0}} \gg$ a Hubble time.  Subsequent interactions between these binaries and single stars greatly reduced $t_{\rm GW}$.  We excluded systems where the initial binary was randomly chosen to contain a BH and NS in our rate estimates to avoid initial condition bias and because these would happen at high redshift and be unobservable.  
\label{fig:tmtm0}}
\end{figure}

%%%%%SECTION 3.1%%%%%
\subsection{Single merger scenario}
\label{sec:sms}
%%%%%SECTION 3.1.1%%%%%
\subsubsection{Without single BHs}
\label{sec:os}

In this section we describe simulations that ended if the binary containing the BH was disrupted or merged. These models did not require any assumptions about the GC's binary population as a whole because they were terminated as soon as BH lost its companion.  We have computed one such simulation in each of our background clusters, and in all of these simulations we assumed $M_{\rm BH} = 7~{\rm M_{\sun}}$.  

In the lowest density cluster considered here, cluster A, only eight BH--NS binaries were formed in our 2000 runs, none of which merged during the $10^{10}$ yr simulations.  In all of the other clusters, however, some fraction of the BH--NS binaries merged during the simulation.  \autoref{fig:esqa} shows the initial semi-major axis and eccentricity of the BH--NS systems formed in simulations using cluster model D.  Also plotted are lines of constant gravitational radiation merger time.  A majority of the systems lie above the right most line, which corresponds to a merger time $10^{10}$ yr.  If these systems were to evolve in isolation, it would take more than a Hubble time for them to merge. However, in the massive GCs considered here subsequent interactions with single stars extracted energy from the binary's orbit and drove some of these systems to high eccentricity and/or small separations, greatly accelerating the merger process.  This mechanism is illustrated in \autoref{fig:tmtm0}.  BH--NS binaries with expected merger times, given their initial eccentricity and semi-major axis, $t_{\rm GW0} > 10^{7}$ yr underwent several interactions with single stars after their formation, which altered the orbital parameters of the binary and reduced the actual amount of time between the formation of the BH--NS binary and the merger of the binary, $t_{\rm GW}$, by several orders of magnitude.  In these simulations, binary--single star encounters not only produced BH--NS binaries, but in some cases enhanced the rate at which these systems merged.  

In other situations, subsequent interactions reduced the merger rate.   In models that used cluster B, many BH--NS binaries were destroyed when a WD exchanged into the binary.  In this relatively low concentration cluster, the density of WDs was much greater than the density of NSs, so multiple encounters with WDs overcame the low probability of a WD exchanging into the BH--NS binary.  This property of low concentration clusters has implications for the formation of mass transfer binaries, and hence X-ray luminosity of such clusters, which we will investigate in a future publication.  At the end of most runs in cluster B, the BH remains in the cluster.  The BH is ejected in 3\% of the runs.  Of those ejected, 42\% are ejected after a BH--NS merger and 58\% are ejected after a super-elastic collision between the binary and a {background} star.  In the majority of runs (63\%), the BH remains in the cluster with a non-NS binary companion.  The BH loses its companion but remains in the cluster in the remaining 34\%.     

This over-density of WDs did not occur in the high concentration clusters, C and D.  In these simulations, if a BH--NS binary was destroyed by an exchange, it was usually another NS exchanging into the system.  The simulations suggest that in the clusters with $n \ge 5\times10^{5}~{\rm pc^{-3}}$, there is a high likelihood that a BH retained by the cluster after the initial phase of self interaction will become part of a BH--NS binary that merges within a Hubble time.  This, coupled with the high post-merger kicks predicted by numerical relativity, suggests that if GC BHs have the same mass distribution as those in the field, BHs will only be retained in massive clusters with $n \le  10^{5}~{\rm pc^{-3}}$.      
\begin{table*}
\centering
 \caption{Globular Cluster Models and Merger Rates\label{tab:runs}}
  \begin{tabular}{@{}cccccccccc@{}}
	\hline
	Cluster & Binary& $f_{b}$ &$M_{\rm BH}$&$f_{\rm BHNS}$& $N_{\rm BHNS}$& $N_{\rm mrg}$&$\bar{t}_{\rm ex}$&$N_{\rm max}$&$ \mathcal{R}_{GC}$\\
	Model&Population&& ($\rm{M_{\odot}}$)&& &&(yr)&&(${\rm yr^{-1}}$)\\
	\hline
\multicolumn{10}{c}{Single merger scenario without single black holes (\autoref{sec:os})}\\
\hline
A & -- & -- & 7 &$4\times10^{-3}$ &  8&0&--&--&0 \\
B & -- & -- & 7 &  0.080  & 202 & 28 & --&--&$1.4 \times 10^{-12}$\\
C & -- & -- & 7 &0.22&  732& 282 & --&--&$1.4 \times 10^{-11}$\\ 
D & -- & -- & 7 &0.34  & 1423 & 465 & -- & -- & $2.4 \times 10^{-11}$\\
\hline
\multicolumn{10}{c}{Single merger scenario with single black holes (\autoref{sec:singles})}\\
\hline
A & FIR & 0.5 & 7 & $5.0 \times 10^{-3}$ &      10 &       0 & $>10^{10} $&       -- & 0 \\
B & FIR & 0.5 & 7 &  0.18 &      470 &       52 & $    3.64 \times 10^{9}$ &       -- & $     2.7 \times 10^{-12}$ \\
C & FIR & 0.5 & 7 &  0.66 &     2568 &     1002 & $    3.31 \times 10^{9}$ &       -- & $     5.1 \times 10^{-11}$ \\
D & FIR & 0.5 & 7 &  0.87 &     4296 &     1626 & $    1.88 \times 10^{9}$ &       -- & $     8.2 \times 10^{-11}$ \\
\hline
\multicolumn{10}{c}{Multiple merger scenario (\autoref{sec:multi})}\\
\hline
A & FIR & 0.2 & 35 & $ 3.0\times10^{-3} $&        6 &        0 & $>10^{10} $&       0 & 0 \\
B & FIR & 0.2 & 35 &  0.14 &      326 &      102 & $    6.65 \times 10^{9}$ &       1 & $     5.3 \times 10^{-12}$ \\
C & FIR & 0.2 & 35 &  0.42 &     1217 &      806 & $    3.98 \times 10^{9}$ &       2 & $     4.1 \times 10^{-11}$ \\
D & FIR & 0.2 & 35 &  0.64 &     2698 &     1635 & $    3.58 \times 10^{9}$ &       5 & $     8.3 \times 10^{-11}$ \\
\vspace{\parskip}\\
A & FIR & 0.5 & 35 & $2.5 \times 10^{-3}$ &       5 &        0& $>10^{10} $&       0 & 0 \\
B & FIR & 0.5 & 35 &  0.15 &      360 &      124 & $    3.14 \times 10^{9}$ &       1 & $     6.4 \times 10^{-12}$ \\
C & FIR & 0.5 & 35 &  0.54 &     1882 &     1185 & $    3.16 \times 10^{9}$ &       3 & $     6.1 \times 10^{-11}$ \\
D & FIR & 0.5 & 35 &  0.84 &     5679 &     3311 & $    2.01 \times 10^{9}$ &       7 & $     1.7 \times 10^{-10}$ \\
\vspace{\parskip}\\
A & FIR & 0.75 & 35 & $1.5 \times 10^{-3}$ &       3 &        0& $>10^{10} $ &       0 & 0 \\
B & FIR & 0.75 & 35 &  0.15 &      377 &      137 & $    4.22 \times 10^{9}$ &       1 & $     7.0 \times 10^{-12}$ \\
C & FIR & 0.75 & 35 &  0.68 &     3072 &     1881 & $    1.89 \times 10^{9}$ &       4 & $     9.8 \times 10^{-11}$ \\
D & FIR & 0.75 & 35 &  0.85 &     8664 &     5266 & $    8.13 \times 10^{8}$ &       9 & $     2.7 \times 10^{-10}$ \\
\vspace{\parskip}\\
A & OPT & 0.2 & 35 & $3.5 \times 10^{-3}$ &       7 &        1 & $>10^{10} $ &       1 & $     5.0 \times 10^{-14}$ \\
B & OPT & 0.2 & 35 &  0.16 &      364 &      146 & $    4.54 \times 10^{9}$ &       2 & $     7.6 \times 10^{-12}$ \\
C & OPT & 0.2 & 35 &  0.57 &     2076 &     1275 & $    3.57 \times 10^{9}$ &       3 & $     6.5 \times 10^{-11}$ \\
D & OPT & 0.2 & 35 &  0.85 &     8258 &     4128 & $    1.84 \times 10^{9}$ &       9 & $     2.1 \times 10^{-10}$ \\
\vspace{\parskip}\\
A & OBS & 0.05 & 35 & $2.0 \times 10^{-3}$ &       4 &        0 & $>10^{10} $ &       0 & 0 \\
B & OBS & 0.05 & 35 &  0.15 &      341 &      143 & $    4.79 \times 10^{9}$ &       1 & $     7.3 \times 10^{-12}$ \\
C & OBS & 0.05 & 35 &  0.41 &     1151 &      737 & $    3.94 \times 10^{9}$ &       2 & $     3.7 \times 10^{-11}$ \\
D & OBS & 0.05 & 35 &  0.54 &     1760 &     1089 & $    4.42 \times 10^{9}$ &       3 & $     5.5 \times 10^{-11}$ \\
\vspace{\parskip}\\
A & OBS & 0.1 & 35 & $5.5 \times 10^{-3}$ &      11 &        1 & $>10^{10} $ &       1 & $     5.0 \times 10^{-14}$ \\
B & OBS & 0.1 & 35 &  0.13 &      319 &      133 & $    4.58 \times 10^{9}$ &       1 & $     6.8 \times 10^{-12}$ \\
C & OBS & 0.1 & 35 &  0.40 &     1089 &      725 & $    4.05 \times 10^{9}$ &       2 & $     3.7 \times 10^{-11}$ \\
D & OBS & 0.1 & 35 &  0.58 &     2054 &     1250 & $    4.17 \times 10^{9}$ &       3 & $     6.3 \times 10^{-11}$ \\
\vspace{\parskip}\\
A & OBS & 0.2 & 35 & $3.0 \times 10^{-3}$ &       6 &        0 & $>10^{10} $ &       0 & 0 \\
B & OBS & 0.2 & 35 &  0.14 &      337 &      113 & $    4.50 \times 10^{9}$ &       1 & $     5.8 \times 10^{-12}$ \\
C & OBS & 0.2 & 35 &  0.43 &     1313 &      818 & $    3.81 \times 10^{9}$ &       3 & $     4.2 \times 10^{-11}$ \\
D & OBS & 0.2 & 35 &  0.67 &     2980 &     1701 & $    3.56 \times 10^{9}$ &       4 & $     8.6 \times 10^{-11}$ \\
\hline
\end{tabular}
\begin{minipage}{140mm}
\medskip
{Properties of the GC models are given in \autoref{tab:gcparms}, the binary populations are described in \autoref{singles}, $f_{b}$ is the binary fraction, $M_{\rm BH}$ is the mass of the BH, $f_{\rm BHNS}$ is the fraction of runs in which a BH--NS binary was produced, $N_{\rm BHNS}$ is the total number of BH--NS binaries formed in the 2000 run simulation, $N_{\rm mrg}$ is the number of BH--NS mergers during the simulation, $\bar{t}_{\rm ex}$ is the mean amount of time it takes for a single BH to exchange back into a binary, $N_{\rm max}$ is the maximum number of BH--NS mergers in a single run, and $ \mathcal{R}_{GC}$ is the BH--NS merger rate.}
\end{minipage}
\end{table*}

The distribution of BH--NS merger times, $t_{\rm mrg}$, for the different GCs models are shown as solid histograms in \autoref{fig:tmdist}.  We note that $t_{\rm mrg} > t_{\rm GW}$ because $t_{\rm mrg}$ includes the time taken to produce the BH--NS binary. Encounters occurred frequently in simulations of the more massive, higher density clusters, so the BH--NS binaries were formed earlier and, as described above, encounters after their formation accelerated the merger process.  For each cluster we calculated the merger rate as $\mathcal{R}_{GC} = N_{mrg}/(N t_{\rm max})$, where $N$ is the total number of binaries simulated and $t_{\rm max} = 10$ Gyr is the length of each run.  In each cluster, $N$ was slightly less than 2000 because we discarded the runs where the initial binary was randomly chosen to contain a BH and NS.  Although these simulations predicted many BH--NS mergers, a substantial faction of the runs ended prematurely because the BH lost its companion.  For the runs in clusters C and D, nearly 40\% of simulations ended for this reason.  In the next section we will examine how the merger rate changes if such single BHs are allowed to interact with and exchange back into a binary.

%%%%%SECTION 3.1.2%%%%%
\subsubsection{With single BHs}
\label{sec:singles}
Here, we describe a set of simulations that allowed BHs that lost their companions as the result of a binary--single interaction to continue their evolution in the cluster.  Again, we have run one such simulation in each of our background cluster models and assumed that $M_{\rm BH} = 7~{\rm M_{\odot}}$.  We note that in these simulations, each BH could only undergo one merger with a NS because with $M_{\rm BH} = 7~{\rm M_{\odot}}$ the post merger kick imparted to the BH exceeded the escape velocity of the cluster.  In all four simulations, we used the FIR binary population with $f_{b} = 0.5$.  

These changes had little effect on simulations of cluster A.  The interaction rate in cluster A is lower than the other clusters, so the initial BH binaries undergo far fewer encounters.  Only 57 of 2000 runs in this simulation produced a single BH, and none of these single BHs exchanged back into a binary.  On the other hand, allowing the single BHs to find new companions increased the number of BH--NS binaries that formed and the number of these systems that merged in clusters B, C, and D.  The average amount of time that it took for a BH  without a companion to exchange into a binary, $\bar{t}_{\rm ex}$, was 3.64, 3.31, and 1.88 Gyrs for clusters B, C, and D, respectively.  In most cases, a single encounter did not result in the BH exchanging into the binary, so $\bar{t}_{\rm ex}$ is longer than the interaction timescale.  When the BH did exchange into the binary, the resulting systems (in all three cluster models) had an average semi-major axis of 2.8 au and average eccentricity of  $0.8$.  Because the interaction cross section for such wide and eccentric binaries is large, the newly formed systems quickly interacted with additional {background} stars.  In the simulations of cluster D, a BH-NS binary was formed in 87\% of the runs, and 93\% of the runs that produced such a binary ended with a BH-NS merger.  A single run could produce multiple BH--NS binaries as the result of numerous exchanges, but only a single BH--NS merger.  {For example, in cluster D, the average number of BH--NS binaries produced in a single run was 2.1, but  in one run the BH had ten different NS companions before merging.}  The average BH-NS merger rate in cluster B nearly doubled  when we allowed single BHs to gain new companions.  In clusters C and D the value more than tripled.

\begin{figure*}
\includegraphics[width=1.0\textwidth]{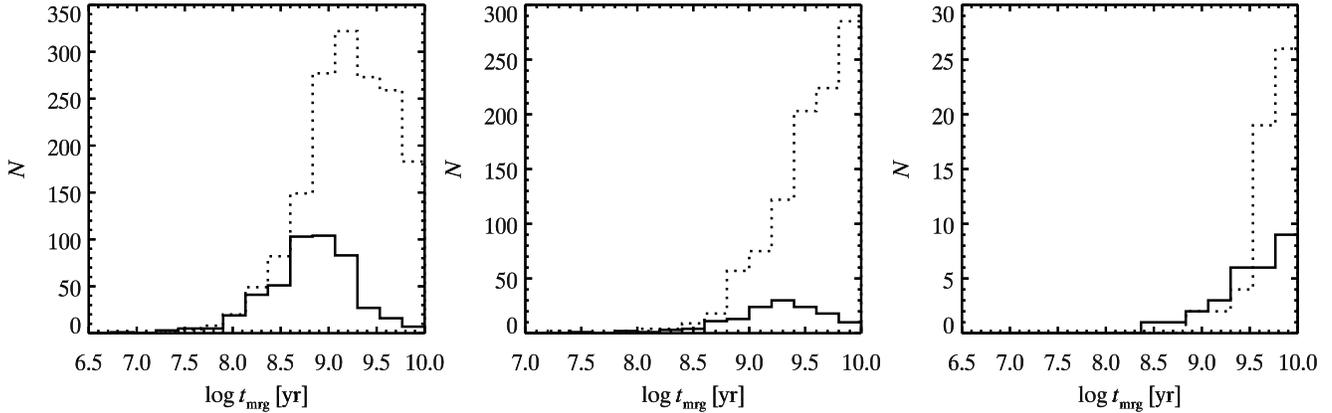}%
 \caption{Distribution of the merger times for BH--NS binaries in clusters D (left panel), C (centre panel), and B (right panel) with $M_{\rm BH} =  7~{\rm M_{\sun}}$. These merger times include the time taken to form the BH--NS system through binary--single star interactions.  The solid histogram is based on runs that extended only until the initial BH binary either merged or was disrupted; it shows the distribution of $t_{\rm mrg}$ when a merger occurred. The dotted histograms show the merger time distribution for runs that allowed BHs that lost their binary companions to exchange into a new binary.  In these runs we have used the FIR binary population with $f_{b} = 0.5$.  With this change, a larger number runs produced a merging BH--NS binary, and the distribution shifts towards longer merger times.  The larger merger times are due to the significant amount of time it takes for the single BH to exchange back into a binary.  In all of the runs shown here, the BH is ejected from the cluster after the merger. \label{fig:tmdist}}
 \end{figure*}

The dotted histograms in \autoref{fig:tmdist} show the distribution of BH--NS merger times in this set of models.  {These dotted histograms are the combination of two $t_{\rm mrg}$ distributions. One distribution, similar to the solid histograms, is due to BHs that merge with a NS before losing their binary companion.  The second distribution is for mergers that occurred after the BH lost its binary companion and had to exchange back into a binary before finally gaining a NS companion and, eventually, merging. For many of these mergers, $t_{\rm mrg}$ is much larger than in the previously discussed cases in which the BH merged  before losing its original companion. Mergers that occurred after the BH became single dominate the $t_{\rm mrg}$ distributions.}  So, in addition to an increased average BH--NS merger rate, these runs also predict that the bulk of these mergers will occur at later times.  However, many of the mergers, especially those in cluster D, still occurred within 3 Gyr of the start of the simulations. This suggests, again, that if GC BHs are of similar mass to those found in the Galaxy and the large post-merger kicks predicted by numerical relativity are correct, most of the BHs that survived the initial phase of self-interaction will be ejected from the cluster after merging with a NS well before the current epoch.  

%%%%%SECTION 3.2%%%%%
\subsection{Multiple merger scenario}
\label{sec:multi}
So far, we have considered scenarios in which the BH receives a post-merger kick that exceeds the escape velocity of the GC.  Under this constraint, each BH can undergo at most one merger.  If GC BHs are more massive than those found in field, then it is possible for the BH to remain bound to the cluster after the merger, regain a NS companion, and merge again. In this section we will discuss models that assumed $M_{\rm BH} = 35~{\rm M_{\odot}}$.  {The magnitude of the recoil velocity is approximately proportional to $q^{2}$, and with a BH mass of $35~{\rm M_{\odot}}$ the recoil velocity becomes comparable to the GC escape velocity.  Thus, we explored the case in which the suppressed kick velocity helped but did not ensure that the post-merger BH would be retained by the cluster. The fraction of post-merger BHs retained in simulations using each background cluster depended on that cluster's escape velocity $v_{esc} \approx \sqrt{2W}\bar{v}_{m}$.  Among all the models of the binary population and varying binary fraction, in simulations using clusters A, B, C, and D, 50\%, 23\%, 20\%, and 10\% of the BHs that underwent a merger were ejected, respectively. Note that only two BH--NS mergers occurred in cluster A, so the ejection fraction derived from the models is smaller than expected.}  However, in these simulations many of the BHs remained bound to the cluster after merging with a NS. A fraction of these single BHs then gained new NS companions and underwent multiple mergers. 

 \begin{figure*}
\includegraphics[width=1.0\textwidth]{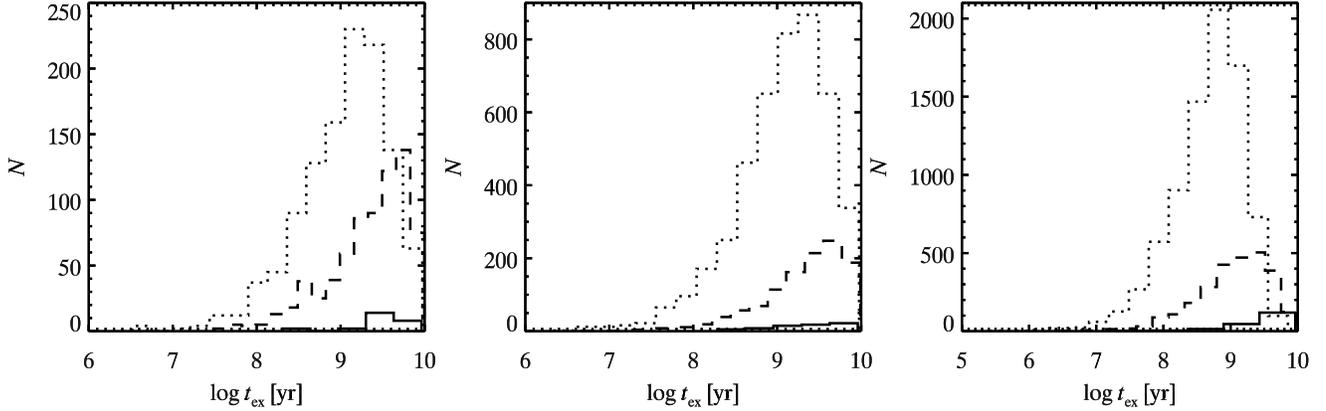}%
 \caption{Distribution of $t_{\rm ex}$, the time for a single BH to exchange into a binary.  Each panel shows the distribution for models using the FIR binary population in cluster B (solid), C (dashed), and D (dotted).  Distributions are show for simulations with $M_{\rm BH} =  7~{\rm M_{\sun}}$ and $f_{b} = 0.5$ (left), $M_{\rm BH} =  35~{\rm M_{\sun}}$ and $f_{b} = 0.5$ (centre), and  $M_{\rm BH} =  35~{\rm M_{\sun}}$ and $f_{b} = 0.75$ (right).  Increasing $M_{\rm BH}$ does not have a large effect on the distribution of $t_{\rm ex}$. Increasing $f_{b}$, on the other hand, results in a narrower distribution that peaks at a lower $t_{\rm ex}$.  Note that the distributions shown in the centre and right panels include $t_{\rm ex}$ for BHs that became single after merging with a NS, while all such systems were promptly ejected from the cluster for the models shown in the left panel.      \label{fig:tedist}}
 \end{figure*}

The number of times a BH can merge with a NS depended on how quickly the single BH could exchange into a binary after each merger.  While the post merger kick was not always large enough to remove the BH from the cluster, the BH was expelled from the core after most mergers.  Before it was able to exchange into a binary, it needed to sink back into the core where the interaction rates were highest.  This process occurred rapidly in all of the simulations.  The mass segregation timescale is proportional to $\bar{m}/M_{\rm BH}$, where $\bar{m}$ is the average stellar mass in the cluster. The BHs in these models were $35~{\rm M_{\odot}}$, which gave $\bar{m}/M_{\rm BH}$ between 0.014--0.027.  Typically, a kicked BH would return to the core on a timescale $O(10^{7})$ yrs.  As can be seen from \autoref{tab:runs}, this timescale is short compared to the total time it takes the single BH to exchange into a binary.  There was, therefore, little difference between the $t_{\rm ex}$ distributions for BHs that became single because of a BH--NS merger and those that lost their companions after a binary--single interaction. We make no distinction between these distributions in the discussion below.   

\begin{figure}
  \includegraphics[width=0.5\textwidth]{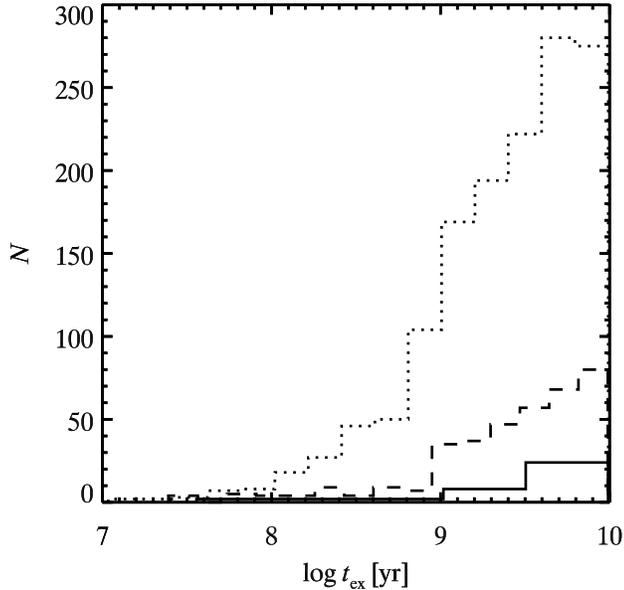}%
  \caption{Distribution of $t_{\rm ex}$ for models in cluster B (solid), C (dashed), and D (dotted).  In these simulations we used the OBS binary population, $M_{\rm BH} =  35~{\rm M_{\sun}}$, and $f_{b} = 0.2$. These simulations tend towards much longer  $t_{\rm ex}$ despite having a similar {\em observable} binary population to the simulations shown in the right panel of \autoref{fig:tedist}. \label{fig:texobs}}
 \end{figure}

\autoref{fig:tedist} shows how the distribution of $t_{\rm ex}$ is affected by the mass of the BH, the structural parameters of the GC, and the binary fraction.  All of the simulations shown here used the FIR binary population.  Changing the mass of the BH had little impact on both the distribution of $t_{\rm ex}$ and its average value.  As expected, $\bar{t}_{\rm ex}$ decreases with increasing cluster concentration.  Increasing the binary fraction produced slightly narrower  $t_{\rm ex}$ distributions with peaks shifted to shorter times.  The $t_{\rm ex}$ distributions for simulations using the OBS binary population with $f_{b} = 0.2$ are shown in \autoref{fig:texobs}.  For each cluster model, the distributions tend towards longer $t_{\rm ex}$ when compared to the models using the FIR population shown in \autoref{fig:tedist}.  This is only a consequence of the larger $f_{b}$ allowed in models that used the FIR population, because the bulk of the FIR binary population is unobservable.  Models that used the FIR population with $f_{b} = 0.2$ had similar or slightly longer $\bar{t}_{\rm ex}$ than models that used the OBS population at the same $f_{b}$.  The difference in composition between the FIR and OBS populations, on its own, did not impact the amount of time it takes a single BH to exchange into a binary.  

Conversely, the composition of the OPT binary population directly impacted $t_{\rm ex}$ and $t_{\rm mrg}$.  Models in cluster D that used the OPT population resulted in an average exchange time that was nearly a factor of two smaller than $\bar{t}_{\rm ex}$ in models that used the OBS or FIR populations with the same $f_{b}$. Furthermore, with the OPT population,  in 27\% of the {encounters that resulted in a} single BH exchanging into a binary, {the BH} exchanged {directly} into a binary with a NS.  Many of these highly eccentric BH--NS binaries quickly merged, and such mergers accounted for 10\% of all mergers in these runs.  In models with the other binary populations, single BHs exchanging directly into a binary with a NS accounted for less than 0.3\% of all mergers.                 
 
 \begin{figure*}
\includegraphics[width=1.0\textwidth]{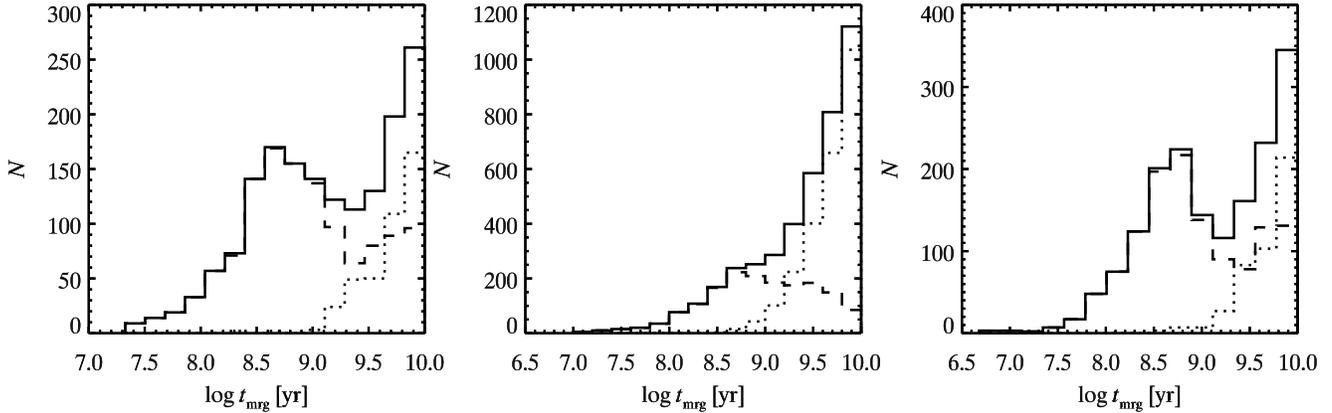}%
 \caption{Merger time distributions for simulations with $f_{b} = 0.2$ using the FIR (left), OPT (centre), and OBS (right) binary population.  Each panel shows the distribution of $t_{\rm mrg}$ for the initial merger during a run (dashed), subsequent mergers during that run (dotted), and the combination of the two (solid).  The initial $t_{\rm mrg}$ distributions for the simulations using the FIR and OBS populations show a clear bimodality.  The peaks above $\log t_{\rm mrg} = 9.5$ in these distributions are due to BHs that lost their binary companions before merging with a NS.  The bimodality is amplified in the total merger time distributions for these runs by the distinct distributions for initial and subsequent mergers.  The $t_{\rm mrg}$ distribution in the OPT case is completely dominated by subsequent mergers.  \label{fig:confb}}
 \end{figure*}
 
 \begin{figure*}
\includegraphics[width=1.0\textwidth]{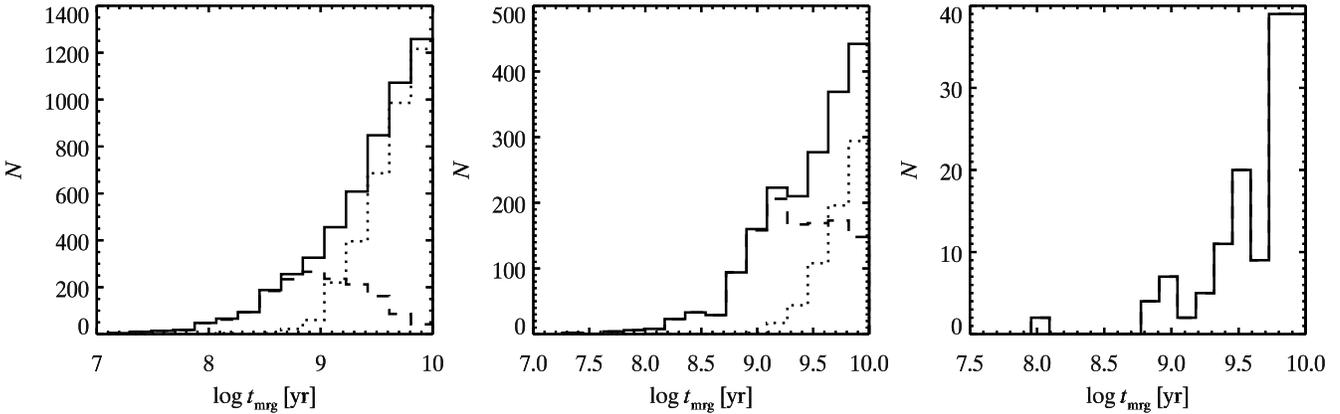}%
 \caption{Distribution of  $t_{\rm mrg}$ for simulations using the FIR binary population with $f_{b} = 0.75$ in cluster D (left), C (centre), and B (right).  Each panel shows the distribution of $t_{\rm mrg}$ for the initial merger during a run (dashed), subsequent mergers during that run (dotted), and the combination of the two (solid).  Subsequent mergers dominate in cluster D.  In cluster B, there are no subsequent mergers.    \label{fig:conpop}}
 \end{figure*}

In fact, the only simulation of cluster B that produced a run with multiple BH-NS mergers used the OPT population.  In all other simulations that used cluster B, the post-merger BHs did not have enough time to exchange into a new binary, gain a NS companion, and merge.  \autoref{fig:confb} illustrates how the assumed binary population affects the merger rate.  Each model shown used $f_{b} = 0.2$, cluster D, and $M_{\rm BH} = 35~{\rm M_{\sun}}$.  As was the case with the $t_{\rm ex}$ distributions, the $t_{\rm mrg}$ distributions for models using the FIR and OBS populations are similar.  In each, the distribution of initial merger times is bimodal.  The first peak corresponds to mergers that occurred before the {initial} BH binary was destroyed in a binary--single encounter.  The second peak, at $t_{\rm mrg} > 3$ Gyr is due to BHs that lost their companion in a three-body interaction and had to re-exchange into a binary system.  The model using the OPT population does not display this bimodality.  The OPT population is tuned to interact with the single BH, so BHs that lose their companions do not remain single for long.  In models using all three populations, the distribution of subsequent merger times peaks in the largest $t_{\rm mrg}$ bin.  This is expected because such mergers cannot occur until after the BH undergone its first merger.  Importantly, in addition to increasing the number of mergers that occurred, and the corresponding merger rate,  $\mathcal{R}_{GC}$, the multiple merger scenario shifted the peak of all $t_{\rm mrg}$ distributions to much later times.  This suggests that such mergers could be observed during the current epoch.

As with the single merger scenario, the more concentrated cluster models produced larger merger rates.  The $t_{\rm mrg} $ distribution for models in clusters B, C, and D are shown in \autoref{fig:conpop}.  Each of the models shown used the FIR binary population with $f_{b} = 0.75$ and $M_{\rm BH} = 35~{\rm M_{\sun}}$. Subsequent mergers dominated the $t_{\rm mrg}$ distribution at $t \ga 1$ Gyr in cluster D.  In the cluster C models, most initial mergers did not happen until $t \ga 1$ Gyr, so the subsequent mergers were not a significant contribution to the total merger rate until late times.  In the cluster B model, none of the BHs underwent multiple mergers, primarily because many of the initial mergers did not occur until the BH had evolved in the cluster for 6 Gyr.    

Clusters with the highest concentrations and binary fractions permitted the largest number of mergers that a single BH can take part in.  The maximum number of BH--NS mergers in a single run, $N_{\rm max}$, is listed for each model in \autoref{tab:runs}.  In a run in cluster D using the FIR binary population and $f_{b}= 0.75$, and in a run in the same cluster model using the OPT binary population with $f_{b} = 0.2$, a single BH merged with 9 NS.  These models also resulted in the largest proportion of runs that resulted in multiple mergers.  In each, 73\% of the BHs that underwent a single merger, underwent at least a second merger.

%%%%%SECTION 4%%%%%
 \section{Discussion} 
 \label{sec:disc}
We have shown that binary--single star interactions in globular clusters can produce BH--NS binaries that will merge within a Hubble time.  The rate at which the mergers occur is sensitive to the GC structural parameters, the mass of the BH, and properties of the GC's binary population.  If the BHs in GCs follow a similar mass distribution to those found in the disk of the Galaxy, then it is unlikely that any high concentration clusters could retain a BH until the current epoch.  In such clusters, most BHs that managed to survive the early period of self-interaction, described in previous studies, would gain a NS companion, undergo a gravitational radiation driven merger, and be ejected from the within $\sim 3$ Gyr.  If, on the other had, GC BHs are more massive than typical field BHs, many clusters could retain a BH and that BH could continue to merge with NSs until late times.               

%%%%%SECTION 4.1%%%%%
\subsection{Uncertainties}
 \label{sec:unc}
The BH--NS merger rates calculated here depend on the uncertain distributions of several quantities.  The distributions of some parameters are better constrained than others  (e.g., $a$ \citealt{Abt:1983} and $e$ \citealt{Duquennoy:1991} distributions, NS retention fraction \citealt{Pfahl:2002}, and the structure of extragalactic GCs \citealt{Strader:2011}).  However, both the GC BH mass function and the nature of GC binary populations strongly influence the predicted rates and are not well constrained by observations.  We have sampled several plausible scenarios for each, and the results suggest that the observed merger rate could be used to probe both the BH mass distribution and binary populations present in GCs.  

Some of the simplifying assumptions that we made in the models also impacted the predicted merger rates.  {In our calculation of the probability that a single BH would exchange into a new binary, we related the densities of single and binary stars using an approximation that diverges for large binary fractions, $f_{b}/(1-f_{b})$.  In the most extreme case considered here, $f_{b} = 0.75$, this term increases the encounter probability by a factor of 3.  {Any overestimate due to this term is of the same magnitude as errors introduced by other approximations made in our calculation.}  In addition to this approximation, in the calculations described above we used a relationship between the rates of specific {\em exchange} encounters to scale the {\em total} encounter rate.  At the very least, this resulted in longer $t_{\rm ex}$ because we only integrated the three-body encounter if the probability for a specific exchange [$(1,2)+{\rm BH} \rightarrow (1,{\rm BH})+2$] was large enough, but we allowed for any outcome of the three-body interaction.}    

We have run two additional simulations to explore how these assumptions affected our calculations.  In these tests, we evolved 500 binaries in cluster D using the FIR binary population with $f_{b} = 0.5$.  In one simulation, marked Retry, we scaled the encounter probability using \autoref{eqn:scale}, as before.  However, in these runs if the probability of exchange was larger than the random number, we re-ran the three-body integrator until the desired exchange, $(1,2)+{\rm BH} \rightarrow (1,{\rm BH})+2$, occurred. If the integration resulted in a different outcome (e.g., flyby or an exchange creating a binary (BH,2)), we changed the orbital phase of the initial binary and the orientation of the encounter, and then simulated the interaction again.  In another simulation, marked Total, we scaled  $P_{\rm (1,BH) + 2}$ by the ratio of the total encounter cross sections, not just the single, desired exchange cross section.  In the mass ratios considered here ($\sim 1:1:35$), the value of $\tilde{\sigma}(X)$ for the BH to exchange into a binary with the primary is roughly equal to $\tilde{\sigma}(X)$ for the BH to exchange into a binary with the secondary, and roughly half the value $\tilde{\sigma}(X)$ for a flyby \citep{Sigurdsson:1993a}.  This results in a factor of 2-3 increase in the encounter probability over the probabilities used in the previous runs.   For the Total runs, as in the runs presented in \autoref{sec:multi}, if an encounter occurred we accepted any outcome of the three-body integration.  

The results of these tests are listed in \autoref{tab:test}.  Both methods for correcting the probability scaling increased the exchange and merger rates.  Furthermore, making the correction either of two ways gave consistent results, illustrating that the probability scaling argument used in our models is valid.  The maximum number of mergers in each of the test runs increased from 7 to 8.  The fact that this quantity did not change by a factor of two, as $\bar{t}_{\rm ex}$ did, suggests that the time it takes a single BH to exchange back into a binary may not be the rate limiting step in the BH-NS merger process.  Finally, these runs suggest that the factor of {a few} uncertainties in the exchange probability resulting from the approximations we made can lead to uncertainties of a similar magnitude in the BH--NS merger rates.  As such, the rates given in \autoref{tab:runs} should be treated as having a factor of $\sim 2$ uncertainty.            

\begin{table}
\centering
 \caption{Probability Scaling Tests\label{tab:test}}
  \begin{tabular}{@{}lccc@{}}
	\hline
	Class &$\bar{t}_{\rm ex}$&$N_{\rm max}$&$ \mathcal{R}_{GC}$\\
	&(yr)&&(yr$^{-1}$)\\
	\hline
 	Original & $2.01\times10^{9}$ & 7 & $1.7\times10^{-10}$\\
	Retry & $9.96\times 10^{8}$&8& $2.3\times 10^{-10}$\\
	Total & $1.18 \times 10^{9}$&8&$2.5\times 10^{-10}$\\
     \end{tabular}
  \end{table}

In addition to uncertainties in the components of the models, physical processes and properties of GCs that we have omitted from the models could alter the BH--NS merger rates.  We have not explicitly included an increase in the binary density at the cluster centre.  Since, next to the BH, NS-containing binaries are the heaviest objects in the cluster they would likely be concentrated near the centre.  The situation could be similar to the models presented here that made use of the OPT binary population, which contained a large fraction of NS binaries.  These models resulted in higher merger rates than models that used OBS or FIR binary populations with the same binary fraction.  On the other hand, a high binary concentration could result in significant heating, expanding the radial profile expected for a population of thermalised binaries.  A high concentration of binaries would also result in binary-binary interactions, which were not included in our models.  Some binary--binary encounters could enhance the merger rate.  For example, the interaction of a BH--MS binary with a NS--MS binary could produce a BH--NS binary.  However, there is a high likelihood that a binary-binary encounter will result in the disruption of a binary. Because the rate of binary-binary encounters increases with $f_{b}$, this limits the binary fraction.  So, binary-binary interactions could cause a reduction in the BH--NS merger rate, compared to the rates predicted here, by limiting $f_{b}$ and making it more difficult for a single BH to find a new binary companion.  Detailed models that include binary-binary interactions are necessary to determine which process dominates.                

%%%%%SECTION 4.2%%%%%
\subsection{LIGO detection rate}
  We estimate the LIGO detection rate following \citet{Belczynski:2007} and \citet{Banerjee:2010}.   We computed the detection rate with
 \begin{equation}
 \mathcal{R}_{LIGO} = \frac{4\pi}{3}\rho_{GC}\left(D_{0}\left(\frac{m_{ch}}{m_{ch,{\rm NSNS}}}\right)^{5/6}\right)^{3}\mathcal{R}_{GC},
 \end{equation}  
 where $\rho_{GC}$ is the space density of GCs, $m_{ch}$ is the chirp mass of the binary ($m_{ch}=(m_{1}m_{2})^{3/5}/(m_{1}+m_{2})^{1/5}$), and $m_{ch,{\rm NSNS}}$ is the chirp mass of a binary with two 1.4 ${\rm M_{\odot}}$ NSs.  {$D_{0}$ is the maximum distance from which the gravitational waves from a merging NS--NS binary can be detected, and has a value of 18 and 300 Mpc for LIGO and Advanced LIGO (AdLIGO), respectively.} Using the galaxy luminosity function parameters estimated in \citet{Croton:2005} and the specific frequency of GCs per $8.5 \times 10^{7}~{\rm L_{\odot}}$ of galaxy luminosity presented in \citet{Brodie:2006}, we calculated $\rho_{GC} = 8 ~{\rm Mpc^{-3}}$.  Our calculations above showed that $\mathcal{R}_{GC}$ is strongly dependent on the cluster mass.   To compute a merger rate that is averaged over all GCs, $\bar{\mathcal{R}}_{GC}$, we used the GC mass function presented in \citet{McLaughlin:1996}.  
 
Although we have shown that the bulk of such mergers occur quickly, and therefore at high redshift, we compute the LIGO detection rate for models with $7~{\rm M_\odot}$ BHs for illustrative purposes.  For these BH--NS binaries, $m_{ch} = 2.6~{\rm M_{\odot}}$.  For the models that do not allow single BHs, we found $\bar{\mathcal{R}}_{GC}= 3.4 \times 10^{-12}~{\rm yr^{-1}~GC^{-1}}$. To make a conservative estimate for the AdLIGO detection rate, we assumed that 10\% of clusters retain one BH and arrived at the rate of $2 \times10^{-3}~{\rm yr^{-1}}$. Using instead the rates from runs that did allow single BHs to exchange back into binaries, we found an AdLIGO detection rate of  $10^{-2}~{\rm yr}^{-1}$.  Even if mergers between $7~{\rm M_{\odot}}$ BHs and NS were occurring in GCs at the current epoch, the event rates are so low that they would not be detected by AdLIGO.  For example, the rates calculated for cluster models B and C may be applicable to massive {intermediate age} clusters.  {With ages of 1-5 Gyr}, these clusters are young enough that any retained BHs would only now be merging with a NS.  However, the space density of {massive intermediate age clusters clusters} is even lower than that of GCs, so the conclusion remains the same.  If GC BHs follow the same mass distribution as field BHs, BH--NS mergers resulting from binary-single star interactions in clusters cannot be detected by AdLIGO.  

The detection rate could be much higher than this conservative estimate if GC BHs are more massive than typical field BHs.  Higher mass BHs enhance the detection rate in two ways.  First, more massive binaries can be detected out to larger distances.  Second, since the post-merger kick is suppressed for high mass ratio inspirals, the BH remains in the cluster and can undergo multiple mergers with NSs.  Models with the FIR binary population and $f_{b} = 0.75$ produced the largest averaged merger rate, $\bar{\mathcal{R}}_{GC} = 4.2 \times 10^{-11} ~{\rm yr^{-1}~GC^{-1}}$. Assuming that 10\% of GCs retain a $35~{\rm M_{\odot}}$ BH, this corresponds to a AdLIGO detection rate of $0.14~{\rm yr^{-1}}$.  If 50\% of GC retain a BH the rate would approach $0.7~{\rm yr^{-1}}$.  The FIR binary population with $f_{b} = 0.75$ represents an optimistic, yet physically motivated population.  On the low merger rate side, models with the OBS population and $f_{b} = 0.05$ would yield an AdLIGO detection rate of $0.04~{\rm yr^{-1}}$ if 10\% of clusters retained a $35~{\rm M_{\odot}}$ BH.  Although the AdLIGO detection rates are not promising, the predicted merger rates are large enough that BH--NS mergers in GCs will be seen in {the next generation of} gravitational wave detectors {(e.g., the Einstein Telescope \citealt{Punturo:2010})}.     

%%%%%SECTION 4.3%%%%%
\subsection{Comparison with previous studies}
Previous studies have estimated the BH--NS merger rate in the field and in GCs.  The field rates are typically calculated using binary population synthesis models that are subject to uncertainties in common envelope evolution, wind mass loss from high mass stars, and natal kicks, resulting in a wide range of predicted AdLIGO detection rates, $0.68-42.8~{\rm yr^{-1}}$\citep{Sipior:2002,Pfahl:2005,Belczynski:2007,OShaughnessy:2010,Belczynski:2010a}.  However, a case study of the likely BH--NS progenitor Cyg X-1 presented in \citet{Belczynski:2011} avoids many of the poorly constrained components of binary population synthesis and predicts much lower field BH--NS merger detection rates in the range $(0.4-2.8)\times10^{-2}~{\rm yr^{-1}}$.   

Our results are consistent with previous studies of the GC merger rate that conclude the cluster merger rate for both NS--NS and BH--NS binaries is much lower than the field rate for corresponding pessimistic, realistic, or optimistic predictions \citep{Phinney:1991,Grindlay:2006,Sadowski:2008}. However, if the uncertainties in the field merger rate push it towards the pessimistic estimates, which is suggested by the models presented in \citet{Belczynski:2011} that circumvent many of these uncertainties,  and multiple mergers increase the GC rate, then the BH--NS mergers detected by AdLIGO might be dominated by, or comprised entirely of, systems that form through multibody interactions in GCs. 

\bigskip
{\noindent ACKNOWLEDGMENTS}

\noindent We thank E.S. Phinney and Francois Foucart for valuable discussion. SS and DFC thank the Kavli Institute for Theoretical Physics for their hospitality. We thank the anonymous referee for helpful comments. DC acknowledges support from the Penn State Academic Computing Fellowship. This material is based upon work supported in part by the National Science Foundation under Grant No. PHYS-1066293 and the hospitality of the Aspen Center for Physics.   

\bibliography{bhns}

\end{document}